\documentstyle[10pt,psfig]{article}
\begin{document}
\begin{center}
{\bf PERFORMANCE OF A SPECKLE INTERFEROMETER }
\end{center}
\vspace{0.3cm}

\begin{center}
S K Saha$^{1}$, A P Jayarajan$^{1}$, G Sudheendra$^{2}$ and A Umesh Chandra$^{2}$\\ 
\end{center}
\vspace{0.3cm}

\noindent
$^{1}$Indian Institute of Astrophysics, Bangalore 560034, India \\ 
$^{2}$Central Manufacturing Technology Institute, Bangalore 560022, India \\ 
\vspace{0.3cm}

\noindent
{\bf Abstract} An interferometer system for use at the 2.34 meter
Vainu Bappu Telescope (VBT), situated at Vainu Bappu Observatory (VBO), Kavalur,     
to obtain speckle-grams of astronomical objects in the visible wavelength,
has been developed. Laboratory tests of resultant intensity distributions
due to a point source, with phase modulation screens, as well as  
the images obtained using this interferometer at the cassegrain end of the said
telescope are discussed.  
\vspace{0.3cm}

\vspace{0.3cm}

\begin{center}
1. {\bf Introduction}
\end{center}
\vspace{0.3cm}

\noindent
Atmospherically induced phase fluctuations distort otherwise flat wavefronts
from distant stars which reach the entrance pupil of a telescope with patches
of random excursions in phase. Such phase distortions restrict the effective
angular resolution of most telescopes to 1 second of arc or worse. Random 
phase excursions are attributed to the refractive index fluctuations created
by pockets of inhomogeneities in the atmosphere characterized by the value of
quasi-coherent areas of diameter r$_{o}$, (r$_{o}$ $\sim$ 10 cm under
general conditions), known as Fried parameter (Fried, 1966). If the exposure
time is shorter ($<$ 20 millisecond) than the evolution time of the phase
inhomogeneities, then each patch of the wavefront would act independently of
the rest of the wavefront resulting in multiple images of the source, called,
'Speckle' (the term 'Speckle' refers to a grainy structure observed when an
uneven surface of an object is illuminated by a coherent source). Its structure
in astronomical images is the result of constructive and destructive
bi-dimensional interferences between rays coming from different zones of 
incident wave. These speckles can occur randomly along any direction within
an angular patch of diameter $\lambda$/r$_{o}$. The sum of several 
statistically uncorrelated speckle patterns from a point source can result
in an uniform patch of light a few seconds of arc wide (conventional image).
\vspace{0.2cm}

\noindent
The technique of speckle interferometry (Labeyrie, 1970) has become an 
invaluable tool for astronomical research in obtaining diffraction limited
spatial Fourier spectrum and image features of the object intensity distribution
from a series of short exposure images through a narrow band filter.
The intensity distribution in the focal plane in case of quasi-monochromatic
incoherent source can be described by the following equations.
\vspace{0.2cm}

\begin{math}
{c(x,y) = {o(x,y) \ast p(x,y)}}
\end{math}
\vspace{0.2cm}

where $c(x,y)$ is the image, $o(x,y)$ is object intensity distribution,
$p(x,y)$ is the telescope-atmosphere point spread function (PSF) and $\ast$ denotes
convolution.
\vspace{0.2cm}

\noindent
The Fourier space relationships between objects and their images are
\vspace{0.2cm}

\begin{math}
{C(u,v) = {O(u,v) \cdot P(u,v)}}
\end{math}
\vspace{0.2cm}

\noindent
Here, $O(u,v)$ is the object spectrum and $P(u,v)$ is the transfer function.
In the conventional speckle interferometry, the ensemble averaged power
spectrum is obtained for a large set of short exposure images.
\vspace{0.2cm}

\begin{math}
{<\mid C(u,v) \mid^{2}> = {\mid O(u,v) \mid^{2} \cdot <\mid P(u,v) \mid^{2}>}}
\end{math}
\vspace{0.2cm}

where $< >$ indicates the ensemble average and $\mid \mid$ the modulus. 
\vspace{0.2cm}

\noindent
The form of transfer function $<\mid P(u,v) \mid^{2}>$ can be obtained
by calculating Weiner spectrum of the instantaneous intensity from the
unresolved star (reference).
\vspace{0.2cm}

\noindent
The 2.34 meter Vainu Bappu Telescope (VBT) at Vainu Bappu Observatory (VBO),
Kavalur can be used extensively to study high resolution features of many
types of celestial objects. These may be in the form of separation and
orientation of close binary stars (separation $<$ 1 arc second), 
shapes of asteroids, sizes of certain types 
of circumstellar envelopes, the structure of active galactic nuclei etc.
The latitude of this observatory gives us access to almost 70$^{o}$ south
of the celestial equator. Most of the observational results beyond 30$^{o}$ south
of zenith at high latitude stations obtained earlier require to be confirmed
utilizing positional advantage of VBT.
This paper presents the design and performance of the new speckle
camera system suitable for operation at the prime focus of this telescope.
\vspace{0.3cm}

\begin{center}
2. {\bf Speckle Interferometer}
\end{center}
\vspace{0.3cm}

\noindent
A speckle interferometer is a high quality diffraction limited camera
where magnified ($\sim$ f/100) short exposure images can be recorded.
Additional element for atmospheric dispersion corrections is necessary
to be incorporated. At increasing zenith distance speckles get elongated 
owing to this effect. Either a pair of Risley prism must be provided for
the corrections or the observation may be carried out using a narrow
bandwidth filter. In this set up, we have used a narrow band filter to
minimize the effect.
\vspace{0.2cm}

\noindent
To arrive at the design of this equipment, preliminary investigations were
made by us with a modest equipment (Saha {\it et al.}, 1987) consisting
of a Barlow lens, a broad band filter in the blue region and a movie
camera. Speckles and interference fringes of various bright stars were 
recorded with the f/13 beam of the 1 meter telescope at VBO. The power 
of this technique was demonstrated by the detection of some telescope
aberrations (Saha {\it et al.}, 1987). Owing to the low quantum efficiency
of photographic emulsion, use of an image intensifier becomes essential
(Breckinridge {\it et al.}, 1979) to record speckle-grams of faint astronomical
objects. New developments in instrumentation technology enable us to detect
the photon events per frame (Blazit {\it et al.}, 1977, Blazit, 1986) up
to a frame rate of $\sim$ 50 Hz. To detect individual photon event,
recording time resolution $\sim$ 1 $\mu$s has been successfully employed
(Papaliolios {\it et al.}, 1985, Durand {\it et al.}, 1987). We have
modified the afore-mentioned set up by replacing movie camera with an EEV 
uncooled intensified CCD (ICCD) camera (385 $\times$ 288) which provides a standard 
CCIR video output (Chinnappan {\it et al.}, 1991) and were able to obtain 
speckle-grams of several close
binaries through a 5 nm filter centered on H$\alpha$ with the cassegrain end
(f/13) beam of the 2.34 meter VBT at VBO. Two of these binaries
were processed using Blind Iterative De-convolution (BID) technique and
reconstructed the Fourier phases (Saha and Venkatakrishnan, 1996).
\vspace{0.2cm}

\noindent
Figure 1 depicts the optical design of the speckle camera system. Provisions
have been made to observe at both the foci (Prime f/3.25 as well as 
cassegrain f/13) of the aforementioned telescope. Before arriving at the
final concept of the design of this interferometer, the optical alignment was
optimized at the laboratory. An artificial star image and the telescope
with f/3.25 beam were generated. Atmospheric seeing was simulated by introducing
various static dielectric cells (SDC) of different sizes etched in glass
plate with hydrofluoric acid. Several glass plates with both regular and
random distribution of SDCs of known sizes have been made and were used 
to produce speckles at the laboratory. The image was magnified to discern
the individual speckles with a microscope objective. A 10 nm filter centered
on OI[5577] was used to reduce the chromatic blurring. For an accurate
evaluation of the performance of the design of this interferometer, we
have also obtained fringes by placing a mask with multiple aperture in
front of the telescope in the laboratory and compared those with the
computer simulation of the intensity distribution (Saha {\it et al.},1988).
The similarity of the observed image shape obtained at the laboratory, as well
as the computer simulated image pattern proved the perfection of the design
of this interferometer.
\vspace{0.2cm}

\noindent
\begin{figure}
\centerline{\psfig{figure=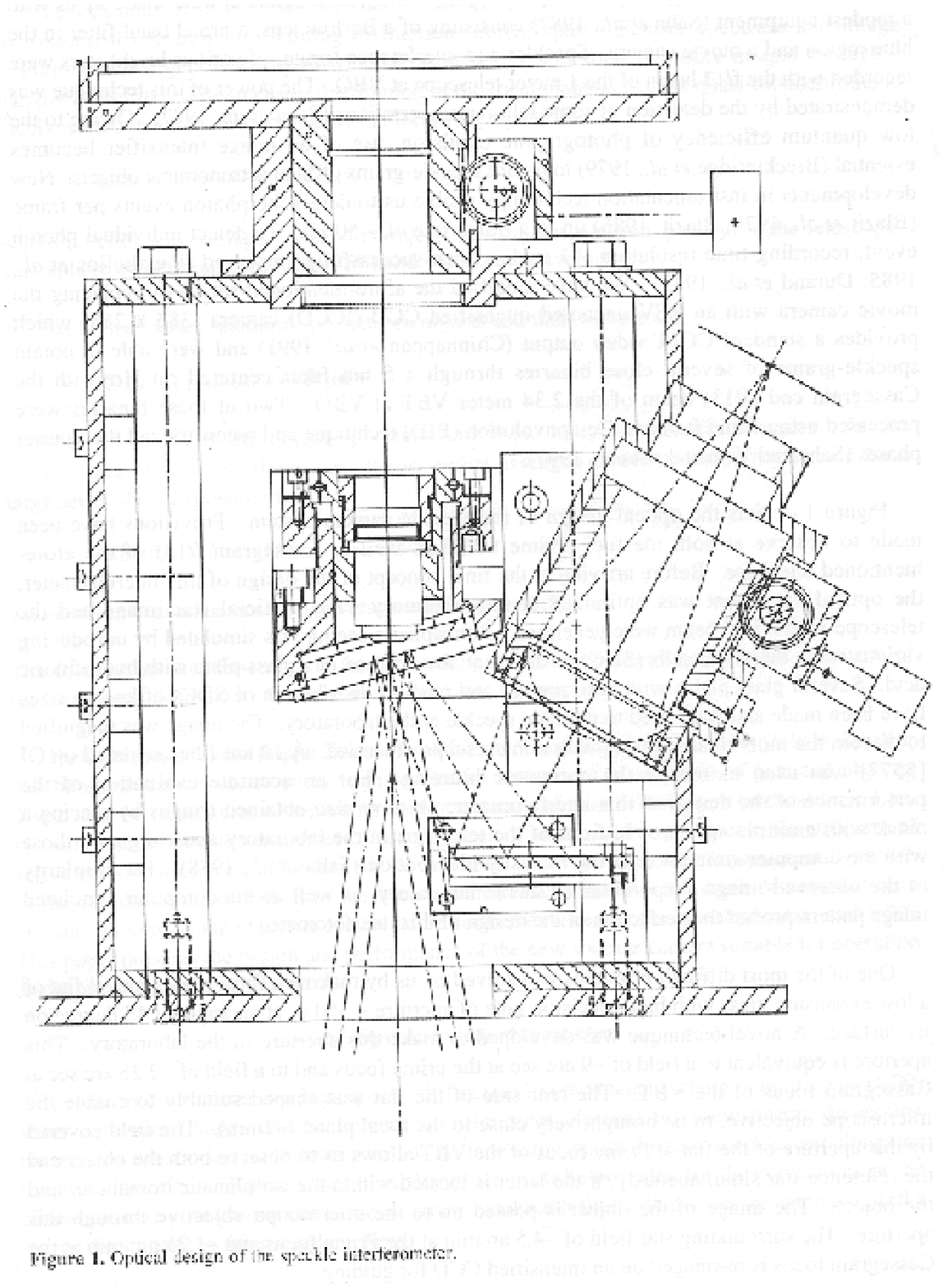,height=16cm,width=12cm}}
\end{figure}

\noindent
One of the most difficult problems was solved by us by making a focal plane
optical flat of a low expansion glass with high precision hole of
aperture $\sim$350$\mu$, at an angle of 15 degree on its surface.
A novel technique was developed to make this aperture in the laboratory.  
This aperture is equivalent to a field of $\sim$9 arc sec at the Prime focus and
to a field of $\sim$2.25 arc sec at Cassegrain focus of the VBT. The rear side
of the flat was shaped suitably to enable the microscope objective, to
be brought very close to the focal plane ($\sim$1mm). The field covered
by this aperture of the flat at Prime focus of the VBT allows
us to observe both the object and the reference star simultaneously,
if the latter is located within the iso-planatic domain around the object.
The image of the object is passed on to the microscope
objective through this aperture. The surrounding star field of $\sim$4.5
arc min. at the Prime focus and $\sim$1.25 arc min. at the Cassegrain focus is
re-imaged on an Intensified CCD for guiding.
\vspace{0.2cm}

\noindent
Before finalizing the mechanical design for the mounts and the housing
of this interferometer, Finite Element Method 
(FEM) analysis was performed to have prior informations about the deflections,
deformation, stress, flexure etc. of the material. The computer simulation
test had shown that the instrument can hold any detector of 10 kgs. weight
kept at a distance of 200 mm away from the rear end of the interferometer
with a flexure of $\sim$1.3$\mu$. The model was analyzed for strength and
deflection for a load of 20 kgs. over the span of the instrument. The analysis
shows a deflection of $\sim$0.7$\mu$. Figure 2 shows the FEM model of the 
structure of speckle interferometer.
\vspace{0.2cm}

\noindent
\begin{figure}
\centerline{\psfig{figure=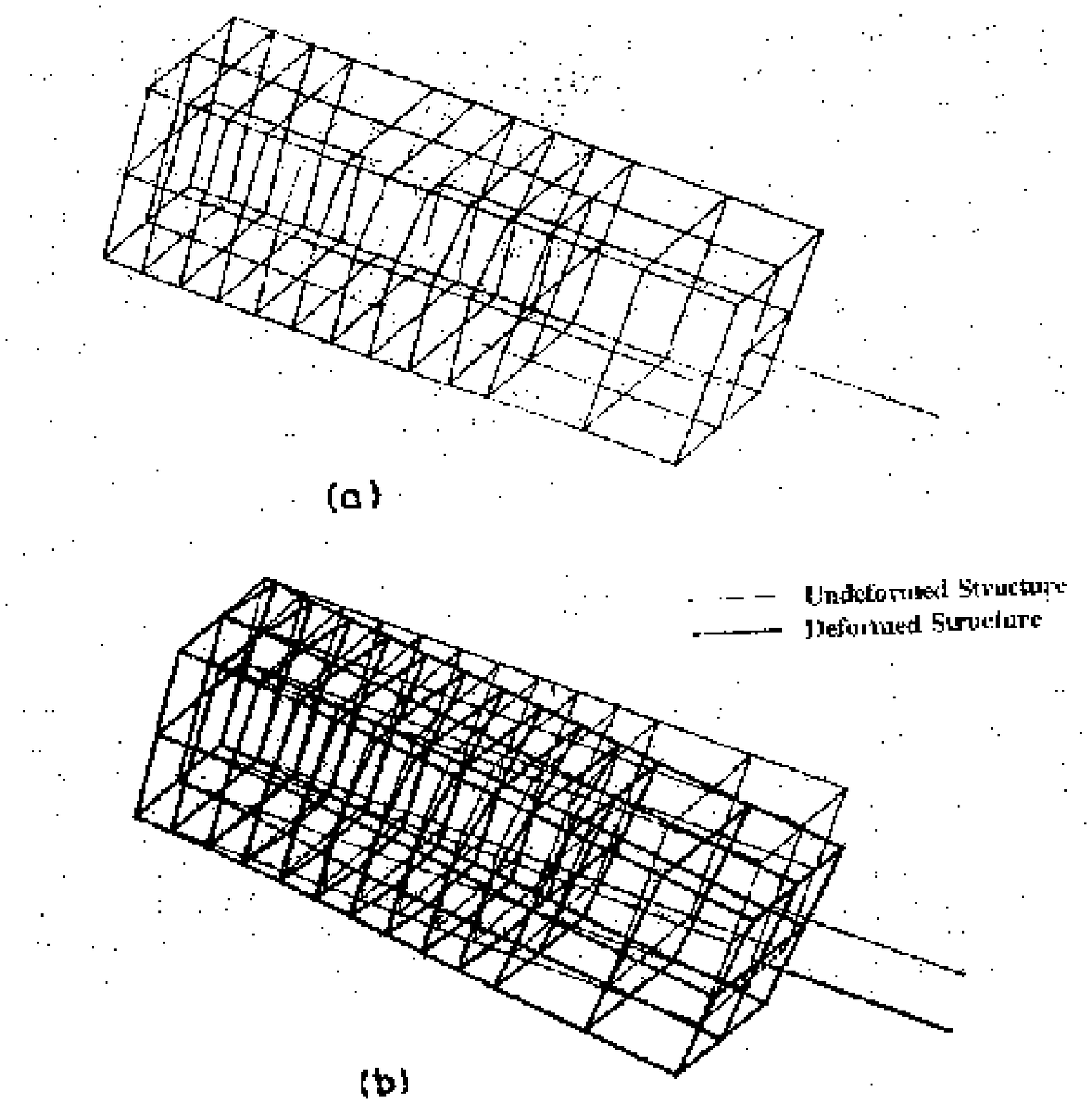,height=14cm,width=12cm}}
\end{figure}

\noindent
The mechanical design requirements which have to be met while designing
the interferometer are:
(i) high accuracy,
(ii) light weight,
(iii) minimum deflection at various orientations of the telescope,
(iv) provision for fine adjustment of the microscope without allowing 
rotation of the image,
(v) provision for fine focusing of the camera without disturbing / rotating
the image,
(vi) provision for adjusting the inclined beam,
(vii) rigid locking of the lens positions during use of the instrument.
\vspace{0.2cm}

\noindent
To assure high accuracy even at varying ambient temperatures, a Martensitic
variety of stainless steel material ( SS 410 ) with low coefficient of 
linear expansion has been used. The instrument has been
conceived as a box made of two end plates of thickness 8 mm and joined
by means of struts of section 22 mm square. The struts have been machined
from a cylinder of 25 mm diameter of required length ground at the ends.
The struts are provided with spigots of 18 mm diameter at ends for 
locating the end plates with corresponding holes. The concentricity
between spigots can be ensured by grinding. The end plates are machined
together and the four holes which receive the spigots as well as the
central hole are machined in one setting on wire cut Electro-discharge
Machining (EDM) to assure required
center distance accuracies. The end plates when locked with the struts,
form a box structure of light weight with required strength to house
the desired mounts with minimum deflection. A bottom plate is clamped on
to two of the struts and this forms a platform on which the various
mounts for the microscope, flat and the lens can be mounted. The mounts
themselves are designed and machined in such a way that lens mounting
holes are at an accurate distance from their bases. In case of any error,
the base can be ground to get the required center height accuracy.
Thus, the plane in which the light travels through the microscope and mirrors
is maintained accurately. The individual mechanisms in each of the mounts
provide for fine adjustment which help in fine focusing of the image. Plate I
shows the photograph of the interferometer.
In order to avoid the reflection from other surfaces, black chrome
plating was done to achieve blackening of the stainless steel.
\vspace{0.3cm}

\begin{center}
3. {\bf Observations}
\end{center}
\vspace{0.3cm}

\noindent
We have successfully made an attempt to observe a few close binary 
systems using this newly built interferometer at the Cassegrain focus
of the VBT. The image scale at the Cassegrain focus (6.7 arc second per mm.)
of this telescope was magnified to 0.67 arc second per mm, using a microscope
objective. This enlarged image was recorded through a 5 nm filter centered on
H$\alpha$ using the ICCD camera (Chinnappan {\it et al.}, 1991). The images
were acquired with exposure times of 20 ms using a Data Translation{$^T$$^M$}
frame-grabber card DT-2861 and subsequently stored on to the hard disk
of a PC486 computer. This computer allowed us to record 64 
images continuously in a few seconds time. Figure 3 shows the speckle-gram
of the $\alpha$-Andromeda. The observing conditions were fair with an 
average seeing of $\sim$2 arcseconds during the nights of 29/30 November 1996.
\vspace{0.2cm}

\noindent
\begin{figure}
\centerline{\psfig{figure=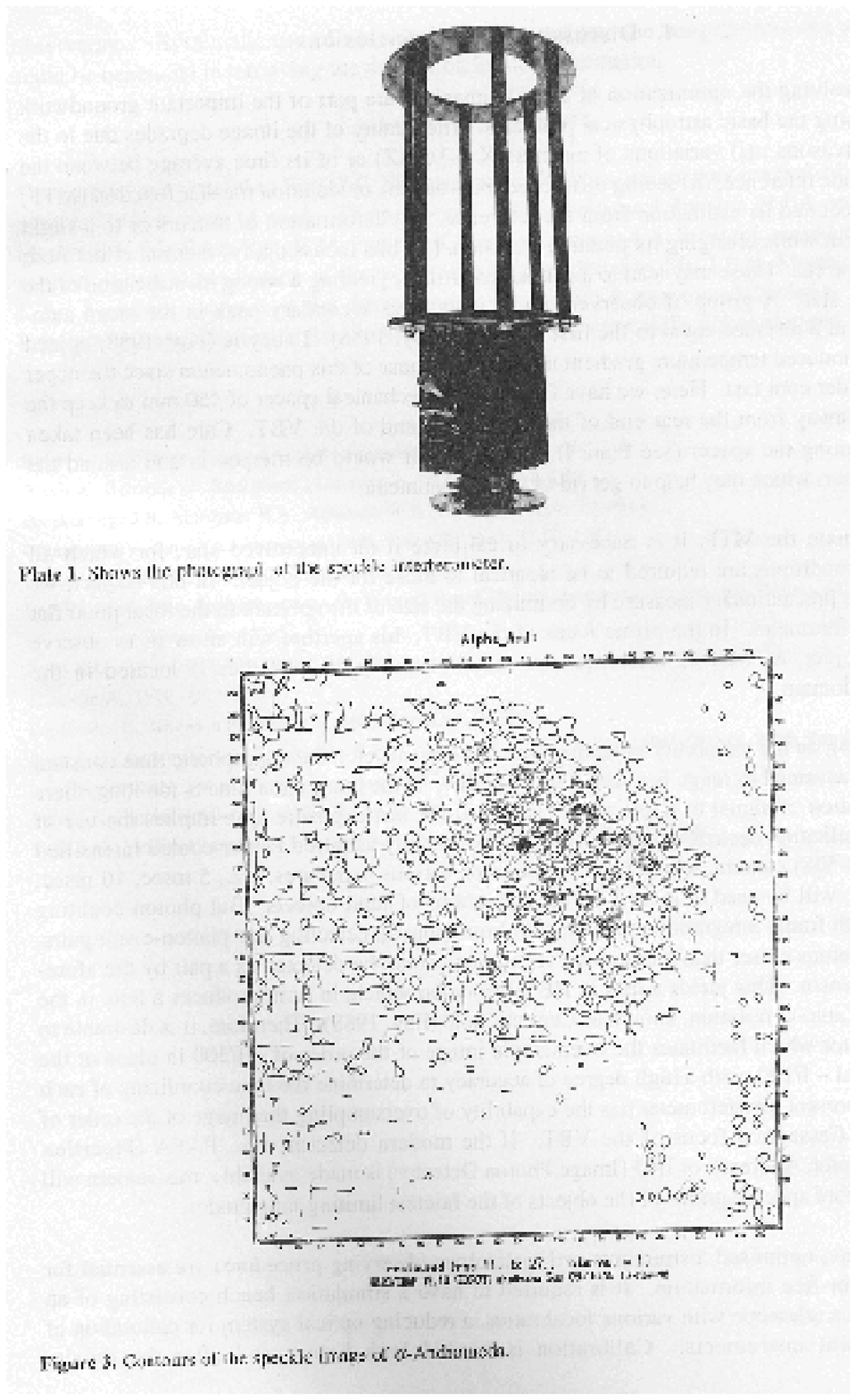,height=16cm,width=12cm}}
\end{figure}

\begin{center}
4. {\bf Discussion and Conclusions}
\end{center}
\vspace{0.3cm}

\noindent
Studies involving the optimization of speckle imaging, are part of the
important groundwork for addressing the basic astrophysical problems. 
The quality of the image degrades due to the following reasons: (i) variations
of air mass X ($\sim$ 1/cosZ) or of its time average between the object 
and the reference, (ii) seeing differences between the modulation transfer 
function (MTF) for the object and its estimation from the reference,
(iii) deformation of mirrors or to a slight misalignment while changing
its pointing direction, (iv) bad focussing, (v) thermal effect from
the telescope etc. These may lead to a dangerous artifact,
yielding a wrong identification of the companion star. A group of
observers found a spurious secondary peak in the mean auto-correlation
at a distance equal to the first Airy ring (Foy, 1988). Labeyrie (Foy, 1988)
opined that spider-induced temperature gradient might be the cause of this
phenomenon since the upper faces of spider cool fast. Here, we have  
fabricated a mechanical spacer of 450 mm to keep the
instrument away from the rear end of the Cassegrain end of the VBT.
Care has been taken while designing the spacer (see Plate I) that no
hot air would be trapped in and around the interferometer, which may   
help to get rid of such phenomena. 
\vspace{0.2cm}

\noindent
To estimate the MTF,
it is neccessary to calibrate it on unresolved star, for which all
observing conditions are required to be identical to those for the object.
In this respact, we have taken a precautionary measure by optimising
the size of the aperture in the focal point flat of this interferometer.
In the prime focus of the VBT, this aperture will allow us to observe both
the object, as well as reference star simultaneously, if the latter is
located in the isoplanatic domain.
\vspace{0.2cm}

\noindent
Depending on the variability in temperature, air currents etc., the
atmospheric time constant is generally assumed to range between 1 and 100
msec. If the integration time is too long, there is a degradation of signal 
to noise ratio. The need for fast exposure times implies the use of detectors
facilitating desired integration time. The recently acquired 
Peltier-cooled Intensified CCD (386 $\times$ 578) camera, which has the option of  
various exposures, viz., 5msec, 10msec, 20msec etc. will be used to record
the speckle-grams of faint objects. But photon-counting detectors with frame 
integration are subject to limitations in detecting fast photon-event pairs. A 
pair of photons closer than a minimum separation cannot be detected as a pair
by the afore-mentioned sensor. This yields a loss in HF information which, 
in turn, produces a hole in the centre of the auto-correlation, known as
Centreur hole (Foy, 1988). Therefore, it is desirable to have a detector
which facilitates the oversample image of the order of $\sim$ f/500 in place
of the former typical $\sim$ f/100, with a high degree of accuracy to
determine the time coordinate of each event. The present interferometer
has the capability of oversampling the image of the order of f/520 at 
the Cassegrain focus of the VBT. If the modern detector, viz., PAPA
(Precision Analogue Photon Address), or IPD (Image Photon Detector) is made
available, this camera will be able to record speckle-grams of the objects
of the faintest limiting magnitudes.
\vspace{0.2cm}

\noindent
To reiterate, optimised instruments and meticulous observing procedures
are essential for obtaining error-free information. It is required to
have a simulation bench consisting of an artificial star, a telescope with
various focal ratios, a reducing optical system for calibration of the focal
point instruments. Calibration is needed both before and after the on-site
observations. Systematic use of simulated image to validate the image processing
algorithms could be beneficial in retrieving the diffraction limited information.
\vspace{0.3cm}

\noindent
{\bf Acknowledgement} The authors are grateful to Prof. J C Bhattacharyya,
for his constant encouragement during the progress of this piece of work. 
The personnel of the mechanical division of IIA, in particular Messrs B R 
Madhava Rao, R M Paulraj, K Sagayanathan and A Selvaraj, provided excellent
support during execution of the work. The help rendered by Mr. J R K Murthy  
of C M T I, for computer analysis of the design and by Dr. Indira Rajagopal
of National Aerospace Laboratory, Bangalore for the black chrome plating
are also gratefully acknowledged.

\newpage
\begin{center}
{\bf References}
\end{center}
\vspace{0.3cm}

\noindent
Blazit, A., 1986, Proc. 'Image detection and quality' - SFO ed SPIE, 
{\bf 702}, 259. \\
\noindent
Blazit, A., Bonneau, D., Koechlin, L. and Labeyrie, A., 1977, Ap. J., 
{\bf 214}, L79. \\
\noindent
Breckinridge, J. B., McAlister, H. A. and Robinson, W. A., 1979, App. Opt.,
{\bf 18}, 1034. \\
\noindent
Chinnappan, V., Saha, S. K. and Faseehana, 1991, Kod. Obs. Bull. {\bf 11}, 87. \\
\noindent
Durand, D., Hardy, E. and Couture, J., 1987, Astron. Soc. Pacific. {\bf 99}, 686 \\
\noindent
Foy, R., 1988, Proc. 'Instrumentation for Ground-Based Optical Astronomy - 
Present and Future', ed. L. Robinson, Springer-Verlag, New York, 345. \\
\noindent
Fried, D. L., 1966, J. Opt. Soc. Amm., {\bf 56}, 1372. \\
\noindent
Labeyrie, A., 1970, Astron. Astrophys., {\bf 6}, 85. \\
\noindent
Papaliolios, C., Nisenson, P. and Ebstein, S., 1985, App. Opt., {\bf 24}, 287. \\
\noindent
Saha, S. K., Jayarajan, A. P., Rangarajan, K. E. and Chatterjee, S., 1988, Proc.
ESO-NOAO conf. 'High Resolution Imaging Interferometry', ed. F. Merkle,
Garching bei Munchen, FRG, 661. \\
\noindent
Saha, S. K. and Venkatakrishnan, P., 1996, Submitted to Bull. Astron. Soc. Ind. \\
\noindent
Saha, S. K., Venkatakrishnan, P., Jayarajan, A. P. and Jayavel, N., 1987,
Curr. Sci., {\bf 57}, 985. \\
\vspace{0.5cm}

\begin{center}
{\bf Figure captions}
\end{center}
\vspace{0.3cm}

\noindent
1. Fig. 1: depicts the optical design of the speckle interferometer.
\vspace{0.2cm}

\noindent
2. Fig. 2: 2a, 2b show the Finite element model of the structure, and the
same with the load of 20 kgs. over the span of the instrument respactively.
\vspace{0.2cm}

\noindent
3. Fig. 3: shows the contours of the speckle image of $\alpha$-Andromeda.
\vspace{0.2cm}

\noindent
Plate I. shows the photograph of the speckle interferometer.
\end{document}